\documentclass[twoside]{dis08}
\usepackage[latin1]{inputenc}
\usepackage[dvips]{graphicx,epsfig,color}
\usepackage{wrapfig,rotating}
\usepackage{amssymb,amsmath,array}

\begin{document}
\title{Deeply Virtual Compton Scattering at HERA and Prospects on Nucleon Tomography}

\author{Laurent Schoeffel
%
\vspace{.3cm}\\
%
CEA Saclay, DAPNIA-SPP, 91191 Gif-sur-Yvette Cedex, France
}

\maketitle

\begin{abstract}
Standard parton distribution functions contain neither information on the
correlations between partons nor on their transverse motion,
then a vital knowledge about the three dimensional 
structure of the nucleon is lost.
Hard exclusive processes, in particular DVCS, are essential reactions to go beyond
this standard picture.
In the following, we examine the most recent data and their implication
on the quarks/gluons imaging (tomography) of the nucleon.
\end{abstract}

\section{Introduction}

Measurements of the deep-inelastic scattering (DIS) of leptons and nucleons, $e+p\to e+X$,
allow the extraction of Parton Distribution Functions (PDFs) which describe
the longitudinal momentum carried by the quarks, anti-quarks and gluons that
make up the fast-moving nucleons. 
These functions have been measured over a wide
kinematic range in the Bjorken scaling variable $x_{Bj}$ and
the photon virtuality $Q^2$.
While PDFs provide crucial input to
perturbative Quantum Chromodynamic (QCD) calculations of processes involving
hadrons, they do not provide a complete picture of the partonic structure of
nucleons. 
In particular, PDFs contain neither information on the
correlations between partons nor on their transverse motion,
then a vital knowledge about the three dimensional 
structure of the nucleon is lost.
Hard exclusive processes, in  which the
nucleon remains intact, have emerged in recent years as prime candidates to complement
this essentially one dimentional picture \cite{lolopic}. 

The simplest exclusive process is the deeply virtual
Compton scattering (DVCS) or exclusive production of real photon, $e + p \rightarrow e + \gamma + p$.
This process is of particular interest as it has both a clear
experimental signature and is calculable in perturbative QCD. 
The DVCS reaction can be regarded as the elastic scattering of the
virtual photon off the proton via a colourless exchange, producing a real photon in the final state  \cite{lolopic,dvcsh1,dvcszeus}. 
In the Bjorken scaling 
regime, 
QCD calculations assume that the exchange involves two partons, having
different longitudinal and transverse momenta, in a colourless
configuration. These unequal momenta or skewing are a consequence of the mass
difference between the incoming virtual photon and the outgoing real
photon. This skewedness effect can
 be interpreted in the context of generalised
parton distributions (GPDs) \cite{freund2,lolo2,buk}. These functions
carry information on both the longitudinal and the
transverse distribution of partons.
The DVCS cross section depends, therefore, on GPDs \cite{freund2,lolo2,buk}.

In the following, we examine the most recent data recorded from the DESY $ep$
collider at HERA and their implication
on the quarks/gluons imaging of the nucleon \cite{dvcsh1,dvcszeus}.

\section{Latest News from the Front}

\begin{figure}[!] 
  \begin{center}
    \includegraphics[width=8.9cm]{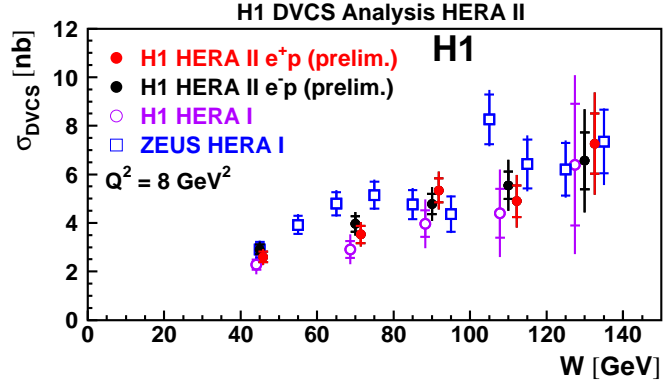}
  \end{center}
  \caption{DVCS cross section for positrons/electrons samples as a function of
$W$.
}
\label{fig1}  
\end{figure}

The first measurements of DVCS cross section have been realised  at HERA within the H1 and
ZEUS experiments \cite{dvcsh1,dvcszeus}. These results are given in the specific kinematic domain
of both experiments,
at low $x_{Bj}$ ($x_{Bj} < 0.01$) but they take advantage of the large range in $Q^2$, offered by the
HERA kinematics, which covers more than 2 orders
of magnitude, from $1$ to $100$ GeV$^2$. It makes possible to study the transition from
the low $Q^2$ non-perturbative region (around $1$ GeV$^2$) towards higher values of $Q^2$ where the higher twists
effects are lowered (above $10$ GeV$^2$).
The last DVCS cross sections as a functon of $W \simeq \sqrt{Q^2/x}$ are presented on figure \ref{fig1}.

\begin{figure}[!]
  \begin{center}
    \includegraphics[width=7cm]{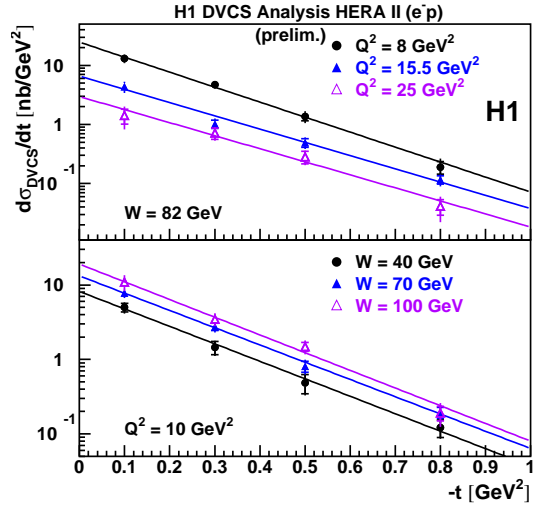}
  \end{center}
  \caption{DVCS cross section, differential in $t$ presented with
              a fit of the form $e^{-b|t|}$. 
}
\label{fig1b}  
\end{figure}

\begin{figure}[!] 
\vspace{-2cm}
  \begin{center}    
    \includegraphics[width=6.5cm]{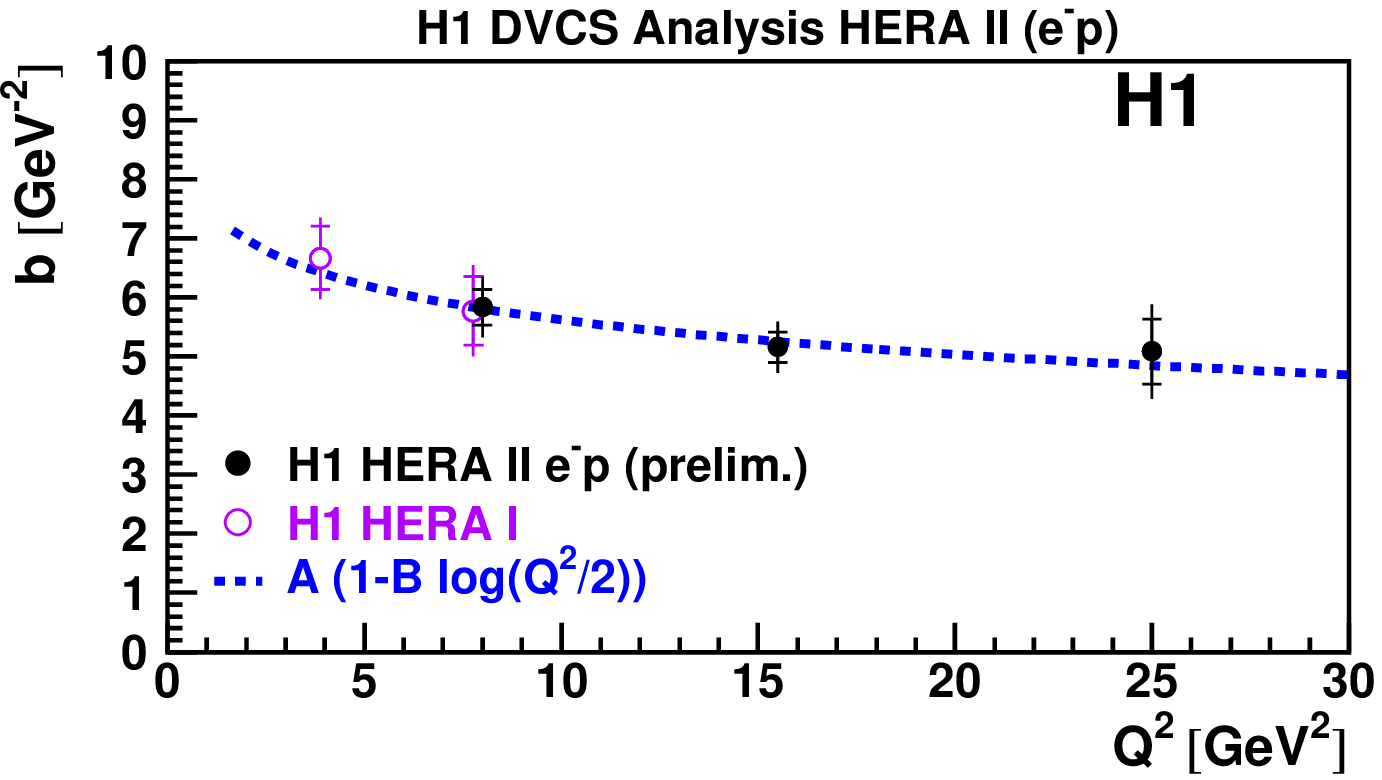}
    \includegraphics[width=6.5cm]{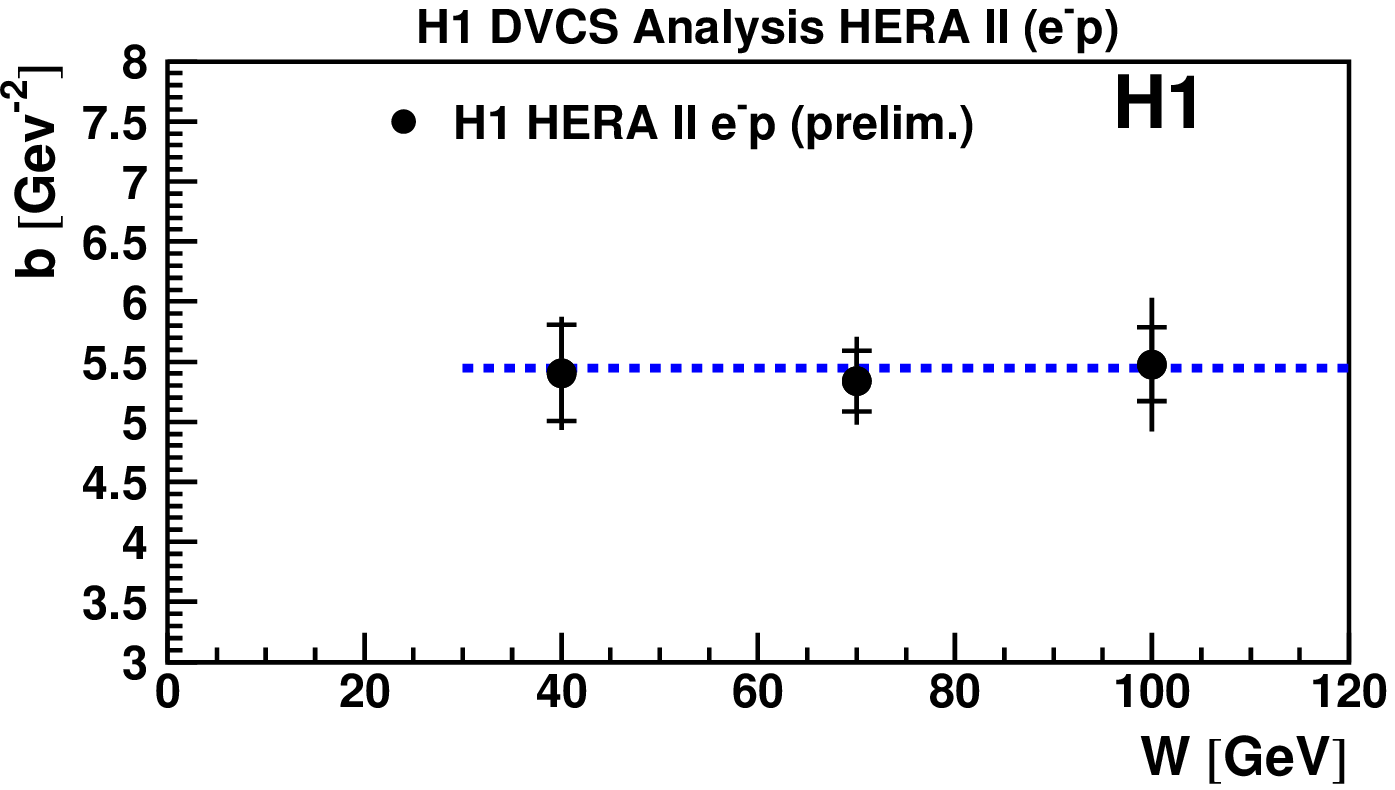}
  \end{center}
  \caption{The logarithmic slope of the $t$ dependence
  for DVCS exclusive production, $b$ as a function of $Q^2$ and $W$, extracted from a fit
  $d\sigma/dt \propto
\exp(-b|t|)$  where $t=(p-p')^2$.
}
\label{fig2}  
\end{figure}

A major experimental achievement of H1 \cite{dvcsh1} has been the measurement of
DVCS cross sections, differential in $t=(p'-p)^2$, the momentum transfer (squared) at the proton vertex.
Some results are presente on figure \ref{fig1b}: we observe  the good description
of $d\sigma_{DVCS}/dt$ by a fit of the form $e^{-b|t|}$. Hence, an extraction of the $t$-slope parameter $b$ is accessible
for different values of $Q^2$ and $W$ (see figure \ref{fig2}).

\section{Nucleon Tomography}

Measurements of the $t$-slope parameters $b$
are key measurements for almost all exclusive processes,
in particular DVCS. Indeed,
a Fourier transform from momentum
to impact parameter space readily shows that the $t$-slope $b$ is related to the
typical transverse distance between the colliding objects \cite{buk}.
At high scale, the $q\bar{q}$ dipole is almost
point-like, and the $t$ dependence of the cross section is given by the transverse extension 
of the gluons (or sea quarks) in the  proton for a given $x_{Bj}$ range.
More precisely, from the  generalised parton distribution $GPD$ defined in the introduction, we can compute
a parton density which also depends on a spatial degree of freedom, the transverse size (or impact parameter), labeled $R_\perp$,
in the proton. Both functions are related by a Fourier transform 
$$
PDF (x, R_\perp; Q^2) 
\;\; \equiv \;\; \int \frac{d^2 \Delta_\perp}{(2 \pi)^2}
\; e^{i ({\Delta}_\perp {R_\perp})}
\; GPD (x, t = -{\Delta}_\perp^2; Q^2).
$$
Thus, the transverse extension $\langle r_T^2 \rangle$
 of gluons (or sea quarks) in the proton can be written as
$$
\langle r_T^2 \rangle
\;\; \equiv \;\; \frac{\int d^2 R_\perp \; PDF(x, R_\perp) \; R_\perp^2}
{\int d^2 R_\perp \; PDF(x, R_\perp)} 
\;\; = \;\; 4 \; \frac{\partial}{\partial t}
\left[ \frac{GPD (x, t)}{GPD (x, 0)} \right]_{t = 0} = 2 b
$$
where $b$ is the exponential $t$-slope.
Measurements of  $b$
presented in figure \ref{fig2}
corresponds to $\sqrt{r_T^2} = 0.65 \pm 0.02$~fm at large scale $Q^2$ for $x_{Bj} < 10^{-2}$.
This value is smaller that the size of a single proton, and, in contrast to hadron-hadron scattering, it does not expand as energy $W$ increases.
This result is consistent with perturbative QCD calculations in terms of a radiation cloud of gluons and quarks
emitted around the incoming virtual photon.

\section{Beam Charge Asymmetry and Conclusions}
The complete parton imaging in the nucleon would need to get  measurements of $b$ for
several values of $x_{Bj}$, from the low $x_{Bj} < 0.01$ till $x_{Bj}>0.1$. Experimentally,
it appears to be impossible. Is it the breakout of quark and gluon imaging in the proton?
In fact, there is one way to recover $x_{Bj}$ and $t$ correlations over the whole $x_{Bj}$
domain: we need to measure a Beam Charge Asymmetry (BCA) \cite{lolopic,freund2}.

\begin{figure}[htbp] 
  \begin{center}
    \includegraphics[width=10cm]{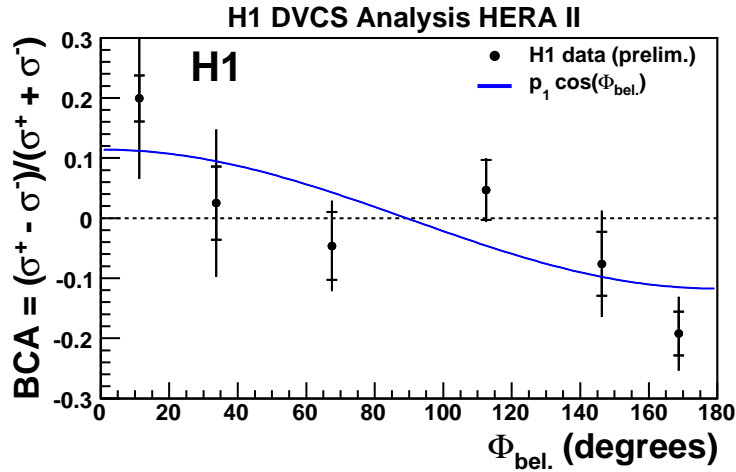}
  \end{center}
  \caption{Beam charge asymmetry as a function of $\phi$ \cite{lolopic}.
}
\label{fig3}  
\end{figure}

The determination of a cross section asymmetry with respect to the beam
charge is realised by measuring the ratio
$(d\sigma^+ -d\sigma^-)/ (d\sigma^+ + d\sigma^-)$ as a function of $\phi$ \cite{lolopic,freund2}.
The lattest experimental result from H1 is presented in figure \ref{fig3} with  a fit in $\cos \phi$.
After applying a deconvolution method to account for the  resolution on $\phi$,
the coefficient of the $\cos \phi$ dependence is found to be $p_1 = 0.17 \pm 0.03 (stat.) \pm 0.05 (sys.)$.

The relative large positve value (more than two standard deviations from zero) is favoring a
non-factorised approach between $x_{Bj}$ and $t$ variables, which means that
on a wide range of $x_{Bj}$, the $t$-slope $b$ is certainly dependent on $x_{Bj}$.

This result represents obviously a major progress in the understanding of the very recent field of the 
parton imaging in the proton. We are at the hedge of the giving a new reading on the most fundamental question to know
how the proton is built up by quarks and gluons...

\begin{footnotesize}




\begin{thebibliography}{99}

\bibitem{url} Slides: \\ 
\verb$http://indico.cern.ch/getFile.py/access?contribId=296&sessionId=31&resId=0&materialId=slides&confId=24657$

\bibitem{lolopic}
  L.~Schoeffel,
  proceedings of 27th International Conference on Physics in Collision (PIC 2007), Annecy, France, 26-29 Jun 2007. 
  arXiv:0707.3706 [hep-ph].


\bibitem{dvcsh1}
F.~D.~Aaron {\it et al.}  [H1 Collaboration],
   Phys.\ Lett.\  B {\bf 659} (2008) 796
  [arXiv:0709.4114 [hep-ex]] ;
  A.~Aktas {\it et al.}  [H1 Collaboration],
   Eur.\ Phys.\ J.\ C {\bf 44}, 1 (2005)
  [hep-ex/0505061] ;
  C.~Adloff {\it et al.}  [H1 Collaboration],
  Phys.\ Lett.\  B {\bf 517} (2001) 47
  [arXiv:hep-ex/0107005].


\bibitem{dvcszeus}
  S.~Chekanov {\it et al.}  [ZEUS Collaboration],
  Phys.\ Lett.\ B {\bf 573}, 46 (2003)
  [hep-ex/0305028].
 
\bibitem{freund2} 
  A.~Freund,
 Phys.\ Rev.\ D {\bf 68} (2003) 096006
  [hep-ph/0306012].

\bibitem{lolo2}
  L.~Schoeffel,
  Phys.\ Lett.\  B {\bf 658} (2007) 33.
  arXiv:0706.3488 [hep-ph].
 
\bibitem{buk} 
  M.~Burkardt,
  Int.\ J.\ Mod.\ Phys.\ A {\bf 18} (2003) 173
  [hep-ph/0207047].



\end{thebibliography}
%

\end{footnotesize}


\end{document}